# Colossal Electric Conductivity in Ag–defect $Ag_5Pb_2O_6$

D. Djurek*, Z. Medunić*, M. Paljević[†] and A. Tonejc[††]

*A. Volta Applied Ceramics (AVAC), Kesten brijeg 5. Remete, HR–10000 Zagreb, Croatia

[†] *Ruđer Bošković Institute, P. O. Box 180, HR–10000 Zagreb, Croatia*

[††] *Phys. Dept., Faculty of Science, University of Zagreb, P. O. Box 331, HR–10000 Zagreb, Croatia*

Byström–Evers compound $Ag_5Pb_2O_6$ has been annealed at 500–540 K under flow of electric current which results in a textured structure and anisotropic colossal electric conductivity ( > $10^9$ $ohm^{-1}cm^{-1}$ ) between 210–525 K. The related physical properties are primarily governed by dissociation of Ag from c-axis channels and lattice strains which in turn depend upon electric current.

In preceding papers [1,2] system Pb–Ag–C–O has been reported as a possible carrier of superconductivity (SC) which extends up to room temperature (RT). More detailed analysis of data appeal for conclusion that Byström–Evers (BE) compound $Ag_5Pb_2O_6$ [3] is responsible for properties reported in [1,2], as well as, for properties listed below appearing in an alternative phase.

Originally $Ag_5Pb_2O_6$ has been prepared in trigonal form by hydrothermal synthesis [3], while Jansen and coworkers [4,5,6] prepared BE compound from $PbO_2$ and $Ag_2O$ in a solid state reaction proposing hexagonal structure shown by Figure 1a. $Ag_5Pb_2O_6$ consists of channels streched along hexagonal c–axis (Figure 1b) and filled with two Ag(1) atoms per unit formula. In order to differ between two types of Ag cations in BE compound more suitable unit formula $Ag_2 \cdot [Ag_3Pb_2O_6]$ may be proposed.

In this work preparation of BE phase $Ag_2 \cdot [Ag_3Pb_2O_6]$ was proceeded from $PbO_2$ and $Ag_2O$ mixed in atomic proportion Ag/Pb=5/2 by the use of magnetic stirrer and benzene. $PbO_2$ contained traces of $H_2O$ visible by infrared (IR) technique but undetectable by conventional thermogravimetric (TG) analysis with sensitivity higher than 0.02 molar percents.

The mixed powder was then fused for 100 hours at 620 K and 200 bar $O_2$. TG decomposition in air to $Pb_3O_4$+Ag at 723 K and subsequent decomposition of $Pb_3O_4$ to PbO at 793 K (inset of Figure 2) revealed the unit formula $Ag_2 \cdot [Ag_3Pb_2O_6]$. X–ray diffraction angles and relative intensities correspond to those cited by Jansen and coworkers.

Heating of BE phase in air, nitrogen or *in vacuo* results in transformations at 550 K and 582 K as it is shown by differential thermal analysis (DTA) diagram (Figure 2) recorded from the pellet of weakly compacted powder (p~30 bar). An annealing at 523 K in air for 48 hours results in a partial release of elementary silver, which is confirmed by X–ray analysis and by visual identification of small silver grains. No traces of $Pb_3O_4$ were recorded which excludes the possibility that silver evolves from the decomposition of BE compound. When a higher uniaxial stress ( about 1500 bar) is applied to BE powder Ag is removed from the compound up to ß > 0.30, with basic hexagonal structure preserved.

The four probe electric resistance was measured in a conventional way by the use of 100 microns gold wire and silver paint. The pellet 5x5x0.4 $mm^3$ was prepared by compression (p~22 bar) of the powder of BE shown by the inset of Figure 3. Heating was performed in air under measuring DC current 100 mA with current contacts connected to A–B and voltage contacts to C–D. The pellet was annealed at 523 K for 24 hours and subsequently cooled to RT. The reheating is followed by downturn of electric resistance at $T_D$= 345 K to less than $10^{-7}$ ohm (Figure 3), which is an ultimate resolution of our DC resistance measurement. This colossal electric conductivity (CEC) state persisted up to 525 K when a partial deterioration appears which manifests as a lower temperature (450 K) of recovery of CEC state by cooling to RT. Based upon repeated preparations the marked chief characteristic of CEC state is the downturn temperature $T_D$ which exclusively depends on strains. After cooling to RT and pulling of the holder with sample on free air CEC state was manifested, under flowing current, next 3 months, independent upon the interchange of electric and voltage contacts. In parallel tests a move of current contacts to A–D and voltage contacts to B–C revealed normal electric resistance of 0.12 ohm which increased in time up to 0.55 ohm, after 3 months of exposure to air and the cited anisotropy also appeared after interchange of B–C (current) and A–D (voltage) contacts. It must be borne in mind that deterioration of CEC state at 525 K has nothing in common with classical transition from superconducting to normal state but it must be looked upon as a result of an appearance of another high temperature structural phase. The cooling below RT results in a reversible deterioration of CEC state near 210 K which is also the lowest $T_D$ recorded by four probe resistance measurement. By heating to RT and switching off the current pellet undergoes in four days a destruction by self–pulverisation. It should be noted that small Ag grains were visible between current contacts and this might be brought in the scheme of mutual dependence of CEC state, Ag dissociation and driven electric current strength. It should be pointed out that anisotropic textured state was also induced by annealing under flowing AC currents (230 cycles per second) supplied from lock–in amplifier.

Part of the pellet, considered by Figure 3 was mounted in a microwave cavity (9,3 GHz) and temperature dependence of the inverse 2Q factor which scales microwave resistance is shown by Figure 4. The sample was firstly cooled down to 90 K and than heated at a rate ~ 1 K/min. Downturn of the microwave resistance is clearly visible at $T_D$ = 210 K , and is in a fair agreement with our four point data, as shown by Figure 3. By repeated cooling–heating cycles conversions to normal and CEC states took place as shown by Figure 4. An application of magnetic field up to 8 T revealed an absence of field dependence of the microwave resistance in CEC state.

An investigation of microwave absorption in modulated magnetic field (MAMMA) pointed out an absence of characteristic absorption hysteresis which contradicts the results reported in [2], and calls for an attention, at least, on two possible modifications of BE samples exhibiting independently classical superconductivity and CEC state.

Anisotropic electric properties induced by electric currents at elevated temperatures suggest an experimental set–up consisting of a metal tube filled with BE powder and we used copper tube with respective outer (OD) and inner (ID) diameters 4 and 2 mm. The tube was filled up to 18 percents of theoretical density of BE phase ( 8,92 g·cm$^{-3}$) and then extruded to OD = 1.5 mm and ID = 0,62 mm. Empty tube of the similar size connected in series with filled one was used as reference sample, and electric contacts were performed by screw tight-fitting. Voltage drops at temperature T on filled and empty tube are denoted as $V_S(T)$ and $V_R(T)$ respectively, whereas the values at RT are labelled as $V_S(0)$ and $V_R(0)$. Then a dimensionless coefficient K = $V_s(T) \cdot V_R(0)/V_R(T) \cdot V_S(0)$ should be plotted against temperature as shown by Figure 5a. It is evident decrease of the resistance of BE filled tube by heating from RT, which resembles a good correspondence to downturn of resistance at $T_D$ = 345 K shown by Figure 3. A net resistance along the wire comes from interruptions in the powdered core as a result of the low filling factor. The filled tube was then replaced by another empty tube of the similar size and corresponding coefficient K was plotted against T and also shown by Figure 5b.

Figure 6 represents: (a) the temperature dependence of the dimensionless factor K for reference sample consisting of full cross-secion copper wire and copper tube filled with BE powder as above, but OD = 0.5 mm and ID = 0.18 mm, and (b) the K factor for the case when the filled tube was replaced by the full cross-section copper wire. It should be noted that voltage contacts were performed by the use of silver paint.

In the next experiment BE filled tube was heated under 0.5 A DC current up to 510 K and then quenched into cold water. Reheating was again performed under 0.5 A and Figure 7 shows temperature dependence of the electric resistance of extruded copper tubes: (a) filled with the insulator $Pb_3O_4$ and (b) filled with BE. It is evident a recovery of the resistance of BE filled tube at 568 K up to the value of that filled with $Pb_3O_4$, and cooling reestablished CEC state at 548 K. At RT resistance was nearly 0.6 value of that before heating, but was partly recovered in time to the initial value. A rather strong noise visible on the voltage contacts usually precedes a complete loss of the resistance above RT in the copper tube filled with BE powder.The electric resistivity calculated from the cross–section and length of the BE core at 523 K is nearly 3 times less than that of silver, although with regard to the low filling factor such a calculation is an overestimate of resistivity of BE phase. In addition, tubes equipped with a concentric copper stud 1 mm in diameter and filled with BE phase in the space between stud and inner tube wall show an absence of CEC when voltage measured on the stud and current contacts attached on the outer tube. This means that CEC is confined to a thin layer of BE phase on the inner wall, that is, conductivity is induced by a tangential interaction of BE phase and electric current flowing in the copper.

In conclusion, it may be abstracted that: [1] BE undergoes transformations at temperatures 550 K and 582 K not reported in preceding literature, [2] annealing at 523 K results in a partial release of silver from c-axis channels, [3] dissociation at 523 K under flowing current results in a textured structure with high anisotropy of electric resistance,[4]there is a close interdependence of strain, dissociation level of Ag and electric current, [5]an absence of magnetic field dependence of microwave resistance data, as well as, an absence of MAMMA effect sound for the alternative appearances to those reported in [2]. Based upon recent experience both effects, observed in BE phase critically depend on strains and related Ag defect.To summarize listed observations, both effects may be recognized, in some instances, as two forms of the same, yet unknown, phenomenon.

An efficient control of the strain and dissociation is a decisive pretext of further research which will shed more light to physical and chemical properties of BE compound.

The authors are indebted to Dr M. Prester from Institute of Physics in Zagreb for an independent evaluation of Figure 6, and to Professor A. Dulčić form the University of Zagreb for microwave resistance measurements. ELKA facility has performed an extrusion of copper tubes.

**References:**

1. D. Djurek, Z.Medunić, M.Paljević, A.Tonejc, Physica C **341-348**, 723 (2000).

2. D. Djurek, Z. Medunić, M. Paljević, A. Tonejc, Physica C **351**, 78 (2001).

3. A. Byström, L. Evers, Acta. Chem. Scand. **4**, 613 (1950).

4. M. Jansen, M. Bortz, K. Heidebrecht, J. Less-Common Met. **161**, 17 (1990).

5. M. Bortz, M. Jansen, H. Hohl, E. Bucher, J. Solid State Chem. **103,** 447 (1993).

6. M. Jansen, M. Bortz, K. Heidebrecht, Z. Kristallogr. 147 (1989).

**Figure captions:**

**Fig. 1.** (a) Unit cell of $Ag_2 \cdot [Ag_3Pb_2O_6]$, (b) c–axis projection of the unit cell of $Ag_2 \cdot [Ag_3Pb_2O_6]$.

**Fig. 2.** Differential thermal analysis (DTA) of $Ag_2 \cdot [Ag_3Pb_2O_6]$ recorded in 1 bar $N_2$. Inset shows the thermogravimetric decomposition curve of $Ag_2 \cdot [Ag_3Pb_2O_6]$ recorded in air.

**Fig. 3.** Temperature dependence of the electric resistance of Ag defect sample $Ag_{2-ß} \cdot [Ag_3Pb_2O_6]$ heated from RT (1) up to 525 K (2), and then cooled to 100 K (3). Inset shows specimen arrangement with electric contacts glued by silver paint.

**Fig. 4.** Temperature dependence of the inverse microwave 2Q factor of the sample from Figure 3. heated from 90 K (1) up to 330 K, then again cooled to 90 K and reheated to 220 K (2).

**Fig. 5.** Temperature dependence of the dimensionless K factor (explanation in text) of the copper tube (OD = 1.5 mm, ID = 0.62 mm) filled with the BE powder (a), and that of the empty copper tube of the similar size (b). Resistance measurement current was I = 0.5 A.

**Fig. 6.** Temperature dependence of the dimensionless K factor (explanation in text) of the copper tube (OD = 0.5 mm, ID = 0.18 mm) filled with the BE powder (a), and a full cross-section copper wire of the similar size (b). Voltage contacts were performed with silver paint. Resistance measurement current was I = 0.5 A.

**Fig. 7.** Temperature dependence of the electric resistance (I = 0.5 A) of the copper tube (OD = 1.5 mm, ID = 0.62 mm) filled with the insulator $Pb_3O_4$ (a), and of the tube of similar size (in series with the former) filled with BE powder (b).

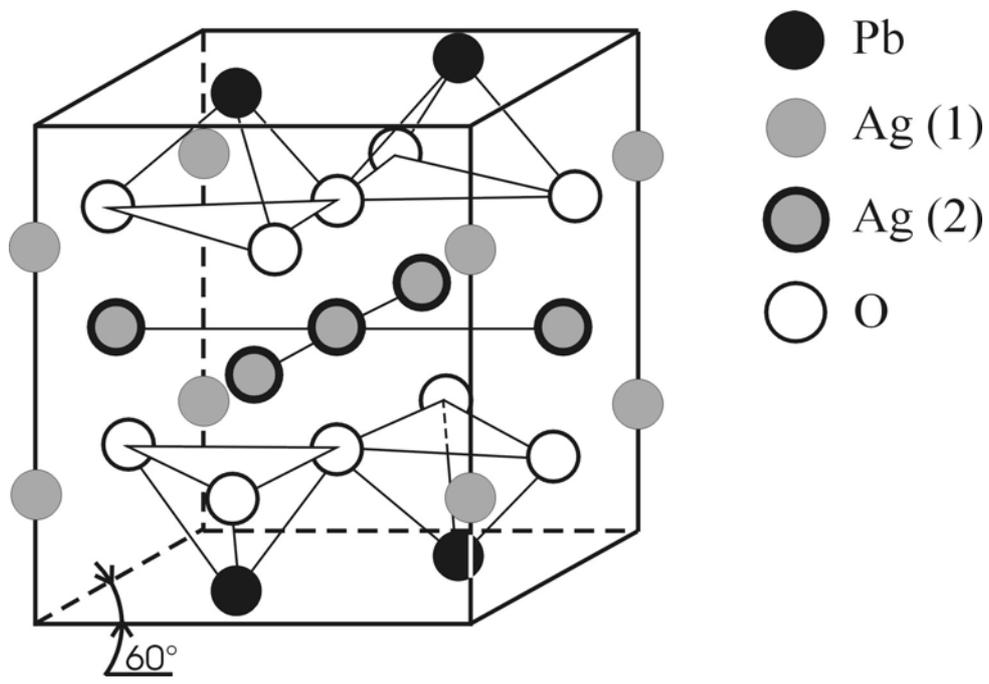

**Fig 1a**

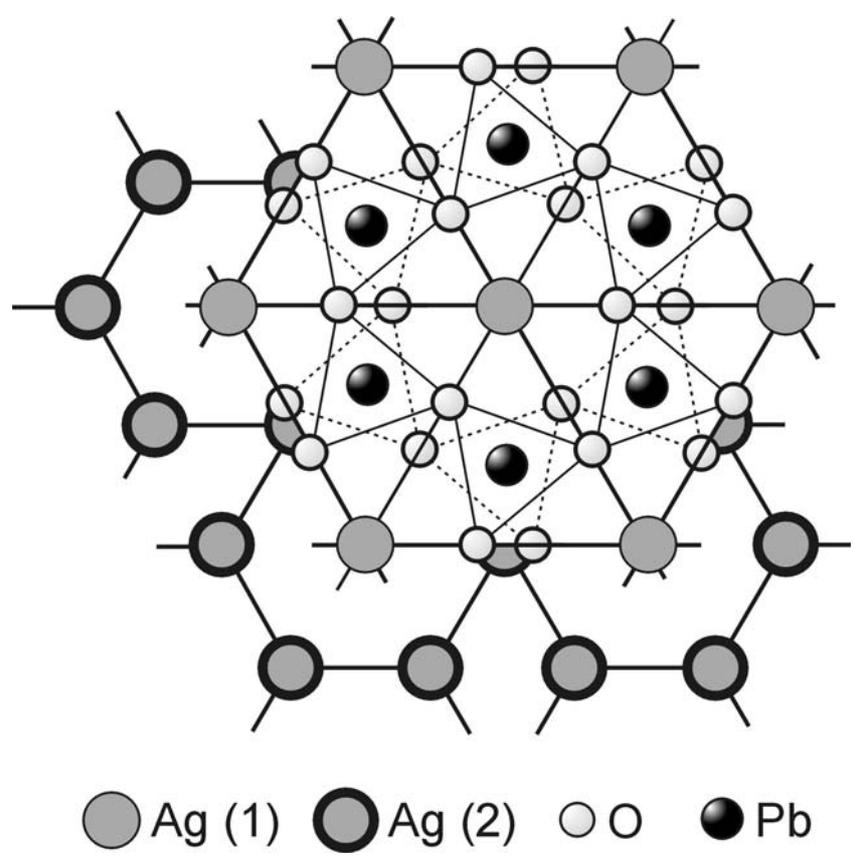

**Fig 1b**

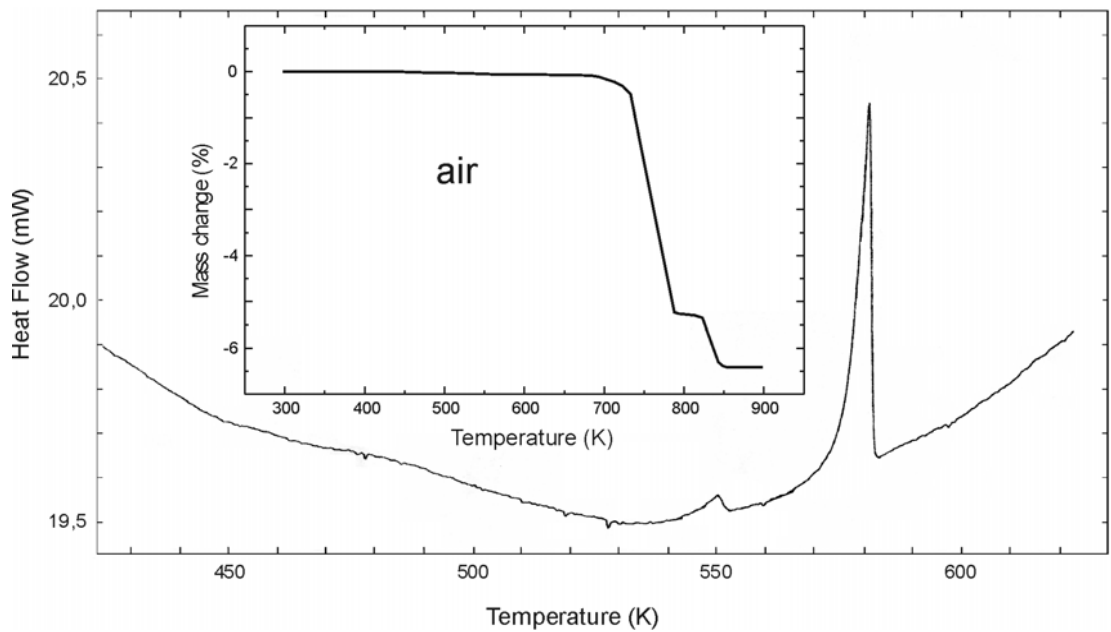

**Fig 2.**

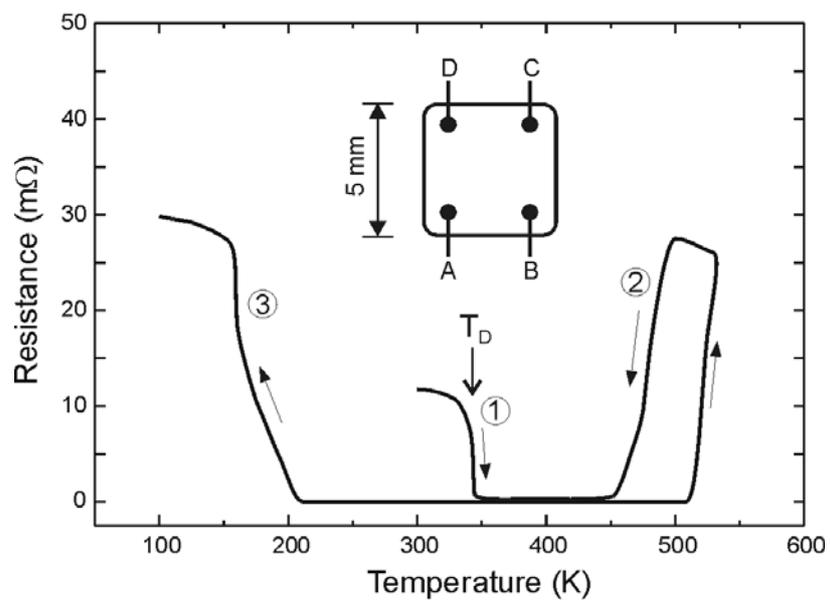

**Fig 3.**

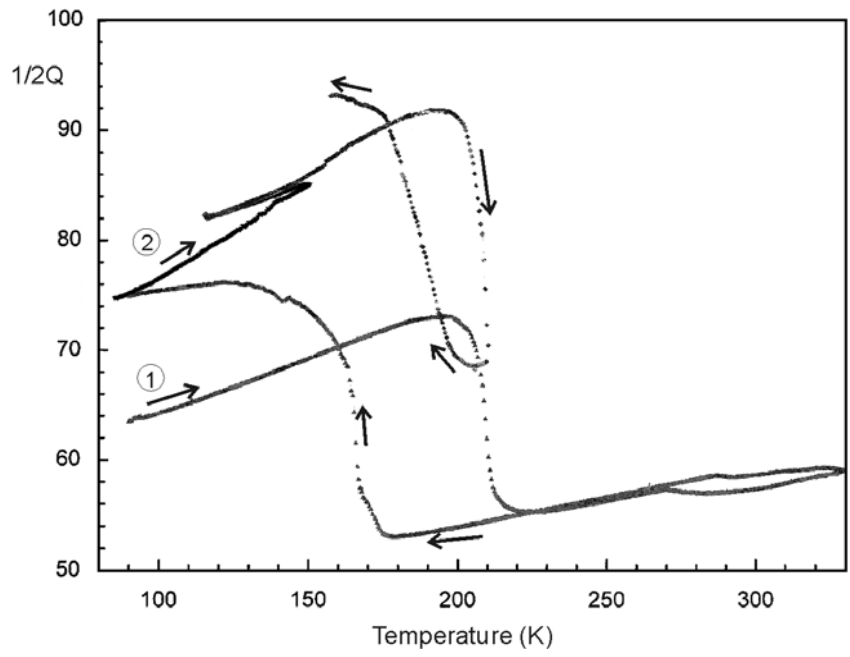

**Fig 4.**

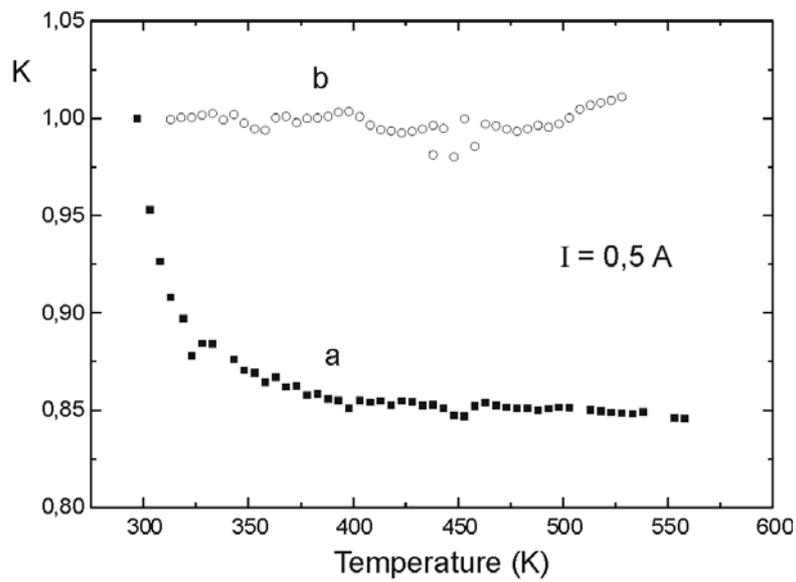

**Fig 5.**

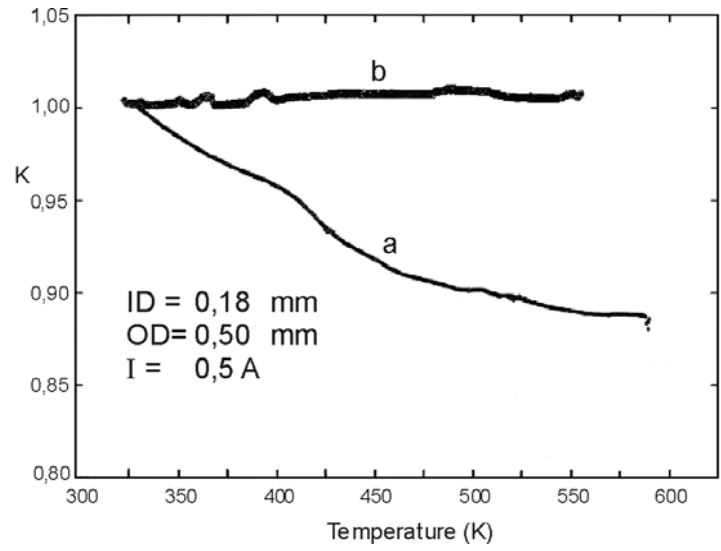

**Fig 6.**

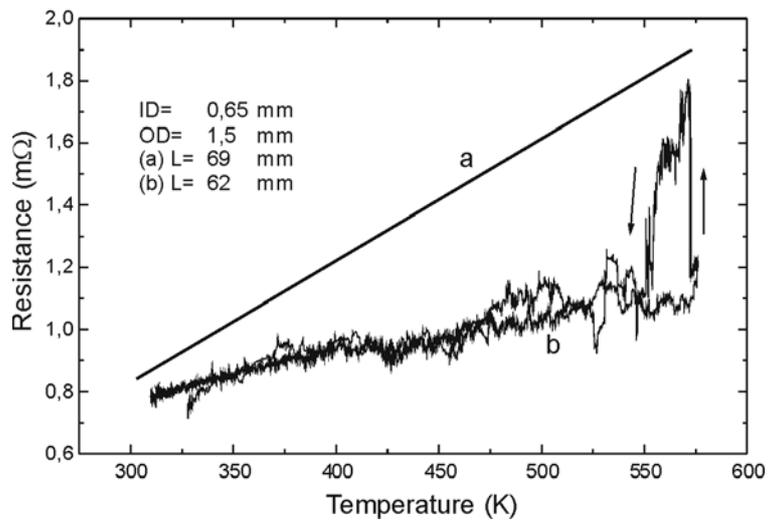

**Fig 7.**